\documentclass[twocolumn,showpacs,preprintnumbers,amsmath,amssymb,superscriptaddress]{revtex4}

\usepackage[dvips]{graphicx}
\usepackage{dcolumn}
\usepackage{bm}
\usepackage[titletoc,title]{appendix}
\usepackage{longtable}
\usepackage{bm}
\usepackage{color}

\begin{document}

\setcounter{LTchunksize}{100}
\setlength\LTcapwidth{6.5in}
\setlength\LTleft{0pt}
\setlength\LTright{0pt}

\title{Microcanonical Ensemble simulation method applied to discrete potential fluids}

\author{Francisco Sastre}
\email{sastre@fisica.ugto.mx}
\affiliation{%
Departamento de Ingenier\'ia F\'isica,\ Divisi\'on de Ciencias e Ingenier\'ias,\\
Campus Le\'on de la Universidad de Guanajuato%
}

\author{Ana Laura Benavides}
\email{alb@fisica.ugto.mx}
\affiliation{
Departamento de Qu\'{\i}mica F\'{\i}sica, Facultad de Ciencias Qu\'{\i}micas,
Universidad Complutense de Madrid, 28040, Madrid, Spain \\
}
\affiliation{%
Departamento de Ingenier\'ia F\'isica,\ Divisi\'on de Ciencias e Ingenier\'ias,\\
Campus Le\'on de la Universidad de Guanajuato%
}

\author{Jos\'e Torres-Arenas}
\email{jtorres@fisica.ugto.mx}
\affiliation{%
Departamento de Ingenier\'ia F\'isica,\ Divisi\'on de Ciencias e Ingenier\'ias,\\
Campus Le\'on de la Universidad de Guanajuato%
}

\author{Alejandro Gil-Villegas}
\email{gil@fisica.ugto.mx}
\affiliation{%
Departamento de Ingenier\'ia F\'isica,\ Divisi\'on de Ciencias e Ingenier\'ias,\\
Campus Le\'on de la Universidad de Guanajuato%
}

\date{\today}

\begin{abstract}

In this work we extend the applicability of the microcanonical ensemble simulation method,
originally proposed to study the Ising model (A.\ H\"uller and M.\ Pleimling,
Int. Journal of Modern Physics C, {\bf13}, 947 (2002)), to the case of simple fluids.
An algorithm is developed by measuring the transition rates probabilities between
macroscopic states, that has as advantage with respect to conventional 
Monte Carlo NVT (MC-NVT) simulations
that a continuous
range of temperatures are covered in a single run.  For a given density,
this new algorithm
provides the inverse temperature, that can be parametrized as a function
of the internal energy, and the isochoric heat capacity
is then evaluated through a numerical derivative.
As an illustrative example we consider a fluid 
composed of particles interacting via a square-well (SW) pair potential of variable range.
Equilibrium internal energies and isochoric heat capacities are obtained
with very high accuracy compared with
data obtained from MC-NVT simulations. These results are important in the
context of the application of H\"uller-Pleimling method to
discrete-potential systems, that are based on a generalization of the SW and Square-Shoulder
fluids properties.

\end{abstract}

\pacs{05.10.-a, 05.20.Jj, 51.30.+i}

\keywords{Microcanonical ensemble, discrete potentials, numerical simulations}

\maketitle

\section{Introduction}

\vspace{0.2cm}

The random cluster model was first introduced by Kasteleyn and Fortuin ~\cite{kas:jpsj69,for:p72}, 
and by Coninglio and Klein ~\cite{con:jpa80} for lattice systems.  
Swendsen and Wang~\cite{Swendsen1987,Newman1999} developed the cluster algorithm
based on the physics of this model and were able to implement
an algorithm using the Widom-Rowlinson and Stillinger-Helfand models for 
fluid mixtures~\cite{Johnson1997,Sun2000},
an example that numerical simulation methods first developed for lattice systems
can also be extended to off-lattice systems.

H\"uller and Pleimling~\cite{Huller2002} have developed a successful cluster algorithm to study 
the Ising model in two and three dimensions within the microcanonical ensemble 
based upon the Broad Histogram Method (BHM)~\cite{Oliveira1996,Oliveira1998,Kastner2000}.
The algorithm works very well because in the Ising model
the energy and magnetization have discretized values.
The H\"uller-Pleimling method (HPM) has two important advantages compared 
with the method proposed by Creutz (CM)~\cite{Creutz1983}: 
First, in a single run a wide range of energies can be covered,
instead of a fixed one as in CM .
Second, the heat capacity can be obtained with a numerical derivative,
whereas for CM it is necessary to calculate the fluctuations
in the Fourier transform of the energy.  
In the CM approach, however, 
no floating-point operations are required and
no random numbers are needed at all if sequential updating is used.

In terms of efficiency the HPM approach compares very well
with the method proposed by Wang and Landau (WLM)~\cite{Wang2001}, that
covers a wide range of temperatures too. 
For example, 
for the 
$32\times 32$ Ising model, 
the average errors for the energy are $0.025\%$ and  $0.035\%$ for HPM and WLM,
respectively, considering the same number 
of Monte Carlo updates  
($7\times 10^5$, 
see reference~\cite{Huller2002}).
  
The first successful attempt to extend BMH 
to continuous systems was made by Mu\~noz and Herrmann~\cite{mun:cpc99}.
In this work we use HPM to study simple fluids, taking advantage
that in this method the entropy is evaluated as
a function of the energy. 
The extension presented here can be applied to discrete-potential (DP) systems,
that also 
have a discrete set of energy values as in the Ising model.
The SW potential is a very simple DP system after the hard-sphere potential,
whose 
structural and thermodynamic quantities have been very well characterized 
along the years 
~\cite{Smith1970,
Henderson1976,ben:mp89,Lonngi1990,
Tang1994,Yuste1994,gil1995,Dawson2001,Reiner2002,Largo2002,Largo2003,
ben:jcp05,ben:jcp06}. 
The SW system
has been extensively applied within
different statistical mechanics approaches~\cite{mac:sm2000,lee:mtnf88}, 
since is usually easier to implement
than other pair potentials in 
statistical-mechanics theories,  
as the Discrete Potential 
and 
Multipolar Discrete Perturbation Theories~\cite{ben:mp99,ben:jcp11},
as well as the Statistical associating fluid theory for chain molecules with attractive potentials of
variable range (SAFT-VR)~\cite{gil:jcp97}.
These and other theories have made possible to
describe phase diagrams of real complex fluids.

The SW potential is a radial potential defined as
\begin{equation}
\phi(r)=\left\{
\begin{array}{ccc}
\infty & \mbox{if} & r\leq \sigma \\
-\epsilon & \mbox{if} & \sigma < r \leq \lambda\sigma \\
0 & \mbox{if} &  r > \lambda\sigma \\
\end{array}
\right.,
\end{equation}
where $r$ is the separation distance between the centers
of two particles; $\sigma$ represents the particle's hard-core
diameter; $\lambda$ is the potential range of an attractive
interaction of depth -$\epsilon$. 

In section II
the method proposed in this work is described in detail,
and its  application to SW systems
is presented in Section III.
Among the thermodynamic quantities that can be calculated, 
we focus our attention on the internal energy and the isochoric heat capacity,
since these quantities are 
more sensitive when perturbation theories are used. As we shall see in this section,
both properties
are predicted
very accurately with  microcanonical simulations
when compared with MC-NVT data.
Finally, in Section IV the main conclusions of this work are given.

\noindent

\section{Simulation Method}

In order to describe the methodology we consider a system of $N$ particles confined
within a volume $V$.
The available energy levels will be labeled as $E_i$, defined as
\begin{equation}
E_i = -i\epsilon = \sum_{k,l\neq k} \phi(r_{kl}),
\end{equation}
where $r_{kl}$ is the distance between the centers of the $k$ and $l$ particles and
$i$ is the number of pair of particles that satisfies $\sigma < r_{kl} \leq
\lambda\sigma$. We denote by
$\Omega(E_i)$ the number of configurations, or microstates, that share the same energy $E_i$.  

A new microstate
is generated when a given mechanism is applied to a previous microstate; for a fluid composed of
spherical particles the mechanism that has to be considered is 
a displacement of a particle, $\Delta\mathbf{r}$. 
When $\Delta\mathbf{r}$ is applied to all $N$ particles for each $\Omega(E_i)$ microstates,
then $N\Omega(E_i)$ new
microstates are created, but only
a small number will have energy $E_j$,
that will be denoted as
$V_{ij}$; in order to satisfy 
the reversibility condition, it is required that $V_{ij}=V_{ji}$,
a relation that holds for both continuous and discontinuous systems,
since the BHM method is an exact theory (see reference~\cite{Poliveira1998}).
The set of movements counted applying BHM is different
to those followed to construct the 
sequence of visited, averaged states, that guarantee equiprobability.
Consequently, it is necessary to consider two different protocols of allowed movements, 
depending on the reversibility or equiprobability conditions to be satisfied.
Then, if a random selection is done of one of the $\Omega(E_i)$ microstates with energy $E_i$
and a new microstate is generated by applying a 
displacement
$\Delta \mathbf{r}$ to a random particle,
then the probability that the system has a new energy $E_j$ is given by 
\begin{equation}
P(E_i \to E_j)=\frac{V_{ij}}{N\Omega(E_i)}\label{probaij}.
\end{equation}
Similarly,
\begin{equation}
P(E_j \to E_i)=\frac{V_{ji}}{N\Omega(E_j)}\label{probaji}.
\end{equation}
Both probabilities can be obtained calculating the rate of attempts $T_{ij}$ to go from 
level $E_i$ to level $E_j$. In order to do this, two variables are required:
\begin{itemize}
\item The number of times that the system spend in level $E_i$, denoted by $z_i$.
\item The number of times that the system attempts to go from level $E_i$ to level $E_j$, 
denoted by $z_{ij}$.
\end{itemize}

Variables $z_i$ and $z_{ij}$  can be evaluated according to the following steps:
\begin{enumerate}
\item With $E_i$ as the initial state, a particle at random is chosen 
and $z_i$ is then redefined as $z_i + 1$.
\item The new energy $E_j$ is evaluated considering that a random displacement is applied to
the previously chosen particle.
\item If $E_j$ is an allowed energy level, then $z_{ij}$ is updated as $z_{ij} + 1$, independently
of the particle's displacement being accepted or not.  The only possible
restriction is the value of the fixed energy chosen to work with, {\em i.e.}
all cases where $E_j<E_{min}$ or $E_j> E_{max}$ are discarded.
\item The probability of an accepted displacement is 1 if $T_{ij}<T_{ji}$, otherwise 
is $T_{ji}/T_{ij}$. 
\end{enumerate}
The last condition assures that all levels are able to be visited with equal probability,
independently of their degeneracy. The initial values for $z_i$ and $z_{ij}$ can be any positive number and 
after a large number of particle's displacement attempts it is observed that
\begin{equation}
\frac{z_{ij}}{z_i}\to T_{ij},
\end{equation}
and the required ratios are given by
\begin{equation}
\frac{T_{ij}}{T_{ji}}=\frac{\Omega(E_j)}{\Omega(E_i)}.\label{dos}
\end{equation}
This algorithm is highly efficient to obtain the ratios $\Omega(E_i)/\Omega(E_j)$
(or the entropy differences $S_i-S_j$, 
according to Boltzmann's relation  $S(E)=k\ln{[\Omega(E)]}$),
since the number of times that the random number generator is used
is smaller than those required in MC-NVT simulations.
The
efficiency of the method increases
if the number of allowed levels $(E_{max}-E_{min})/\epsilon$ is decreased.

Since the 
inverse temperature $\beta(E) = 1/kT$ is obtained from the entropy by deriving it with
respect to the energy, 
\begin{equation}
\beta(E)=\partial S/\partial E 
\end{equation}
then it 
is convenient 
to express the entropy as a series expansion in $\beta(E_i)$,
\begin{equation}
S(E_j)=S(E_i)+\epsilon\eta \beta(E_i)+\ldots,
\end{equation}
where $\eta$
is an integer such that $E_j=E_i+\eta\epsilon$. The rest of the terms in the expansion
can be discarded as long as $N$ is
large enough, obtaining
\begin{equation}
\ln(T_{ij}/T_{ji})\approx \frac{\epsilon\eta}{k} ~
\left.\frac{\partial S}{\partial E}\right|_i. \label{sol_beta}
\end{equation}
This equation can be used to obtain the inverse temperature as a function of the internal energy.

In the Ising model
the energy changes are well defined since they
depend on the number of nearest neighbors in each lattice site and this number is fixed, and the
HPM algorithm can be applied straightforwardly. In the next section we
will explain how this method can also be valid for the case of a SW fluid.

\section{The SW fluid case}

\begin{figure}
\begin{center}
\includegraphics[width=7.50cm,clip]{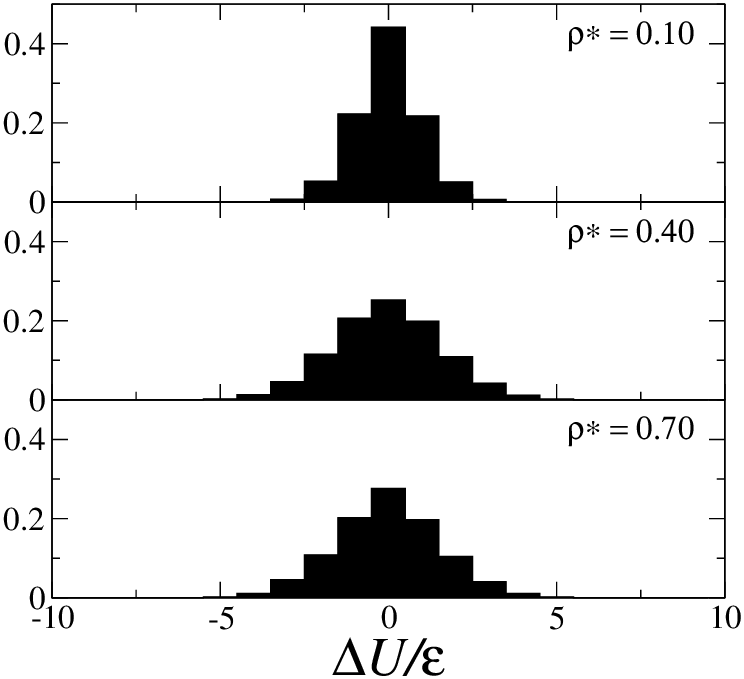}
\caption{\label{histograma}
Histograms for the energy displacements $\eta =\Delta U/\epsilon$
for the SW fluid with attractive range $\lambda=1.5$ and densities $\rho^*=0.1$,
$0.4$ and $0.7$, from top to bottom.
}
\end{center}
\end{figure}

We consider SW systems with attractive ranges $\lambda$= 1.1, 1.3 and 1.5, that are typical values
required to describe molecular and complex fluids. 
Previous to any calculation we must establish the allowed values of $\eta$ for the SW potential.
In the original work by H\"uller and Pleimling this is not necessary
since for the Ising model the energy changes are well defined 
and depend
on the number of nearest neighbors, which is a fixed quantity. For the SW fluid case, 
however,
different jumps of energy 
(i.e., different $\eta$ values ) will occur, with different frequencies among them.
Nevertheless, the values of  $\eta$ can be chosen by considering the more frequent energy jumps
performed by the system. 
This can be achieved using a simple NVT Monte Carlo simulation for several values of 
density and $\lambda$ in the limit of infinity temperature,
a limit in which all energy changes are accepted.
Then histograms can be obtained considering the relative frequency of the energy changes,
as presented in Figure
\ref{histograma} for three different densities of the SW fluid with $\lambda=1.5$.
Applying this method for the densities and SW ranges values  
used in this work, accurate results can be obtained for
the inverse temperature $\beta(E)$ when
$\eta=\pm 1,~\pm 2$
and $\pm 3$.
If we restrict the analysis with  $\eta=\pm 1$ it is possible to obtain reliable values 
of $\beta(E)$ but at the cost 
of increasing
the computing time of the simulations. In any case, it is necessary to evaluate
proper values for $\eta$ each time that a SW system is simulated using HPM. 

Once that $\eta$ values have been obtained, the inverse temperatures are given by 
\begin{equation}
\beta(E_i)=\frac{1}{6}\sum_{\eta=-3}^{3} \frac{1}{\eta}\ln(T_{i,i+\eta}/T_{i+\eta,i}),~~\eta\neq 0.
\end{equation}
Furthermore, from the curves $\beta(E)=S'(E)$ the isochoric heat capacity $c(E)$ can also be obtained 
performing a second derivative,
\begin{equation}
c(E)= -\frac{[\beta(E)]^2}{S''(E)}.
\end{equation}

In order to illustrate this method, 
computer simulations for the SW fluid were performed using a unitary box with periodic boundary conditions, 
considering $N=512$
particles and reduced densities $\rho^* = \rho\sigma^3$ between $0.1$ and $0.8$. 
In all the cases, the reduced 
energy $u^*=E/N\epsilon$ values are restricted between $u^*_{min}$ and $u^*_{max}$
in the supercritical region.
The number of particle's displacement attempts
considered were from $N\times 10^7$, for the smaller energy intervals,
to $2.5\times N\times 10^7$ for the higher energy
intervals and performing 8 different independent runs.

\begin{figure}
\begin{center}
\includegraphics[width=7.50cm,clip]{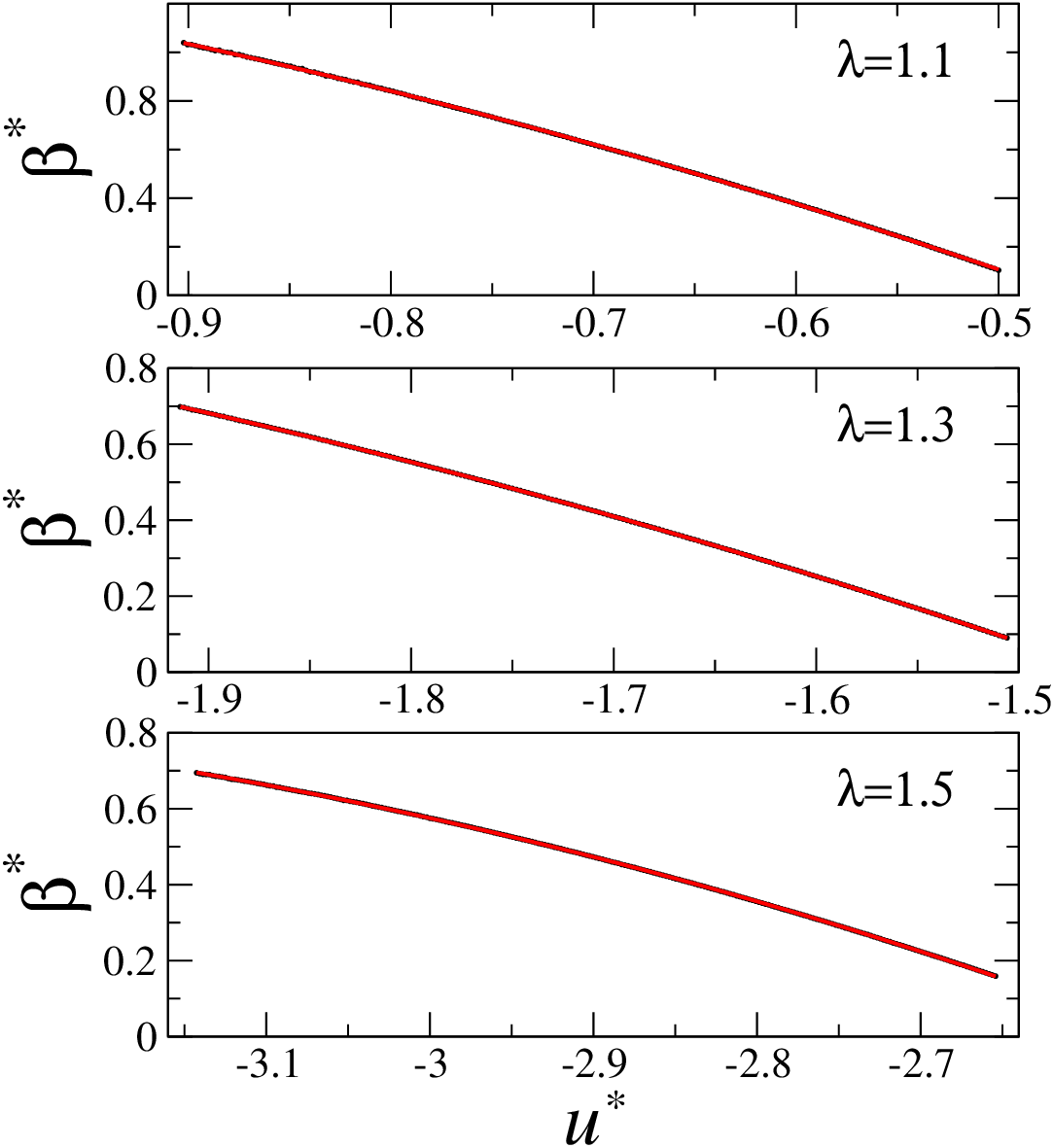}
\caption{\label{ejemplo} (Color on line)
Inverse temperature as a function of energy $u^*$ for density $\rho^*=0.4$ for the SW fluid
with attractive ranges $\lambda=1.1$, $1.3$
and $1.5$. Black dots denote the computer simulation data obtained in this work, and the red solid line
is a second-order polynomial fit.
}
\end{center}
\end{figure}

The inverse temperature obtained values $\beta^*=\epsilon/kT$ 
 were fitted to a second-order degree polynomial on $u^*$
\begin{equation}
\beta^*= a_0+a_1 u^* + a_2 {u^*}^2, \label{polinomio}
\end{equation}
The coefficients obtained from these fitted expressions
are given in tables I-III for the values of density and attractive ranges
used in this work, as well as 
the range of validity
for the fitted expression.
In Figure \ref{ejemplo} we present results for $\rho^*=0.4$ and three different SW systems.
\begin{center}
\begin{table}[ht]
\caption{\label{coef1.1}
Second-order polynomial fit coefficients for $\lambda=1.1$.
}
\begin{tabular}{ccccccccccc}
\hline
\hline
$\rho^*$ & & $a_0$ & & $a_1$ & & $a_2$ & & $u^*_{min}$ & & $u^*_{max}$ \\
\hline
0.10 & &  -0.96087 & &  -14.54988 & &  -23.13050 & & -0.225 & & -0.008 \\
0.20 & &  -1.17362 & &   -7.82579 & &   -6.07659 & & -0.410 & & -0.200 \\
0.30 & &  -1.35121 & &   -5.20500 & &   -2.33237 & & -0.700 & & -0.320 \\
0.40 & &  -1.62688 & &   -4.10229 & &   -1.27205 & & -0.910 & & -0.495 \\
0.50 & &  -1.88153 & &   -3.25117 & &   -0.65977 & & -1.200 & & -0.700 \\
0.60 & &  -2.25890 & &   -2.78357 & &   -0.39227 & & -1.520 & & -0.980 \\
0.70 & &  -2.86484 & &   -2.59557 & &   -0.27284 & & -1.880 & & -1.380 \\   
0.80 & &  -3.27924 & &   -2.10602 & &   -0.09942 & & -2.350 & & -1.750 \\   
\hline
\hline
\end{tabular}
\end{table}
\end{center}
\begin{center}
\begin{table}[ht]
\caption{\label{coef1.3}
Second-order polynomial fit coefficients for $\lambda=1.3$.
}
\begin{tabular}{ccccccccccc}
\hline
\hline
$\rho^*$ & & $a_0$ & & $a_1$ & & $a_2$ & & $u^*_{min}$ & & $u^*_{max}$ \\
\hline
0.10 & &  -1.40800 & &  -6.26420 & &  -4.01083 & &  -0.550 & &  -0.300 \\
0.20 & &  -2.01277 & &  -4.20715 & &  -1.44496 & &  -1.000 & &  -0.650 \\
0.30 & &  -2.89267 & &  -3.80293 & &  -0.90423 & &  -1.430 & &  -1.050 \\
0.40 & &  -4.26284 & &  -3.99128 & &  -0.73094 & &  -1.920 & &  -1.500 \\
0.50 & &  -6.41629 & &  -4.57759 & &  -0.67761 & &  -2.450 & &  -2.000 \\
0.60 & &  -9.80357 & &  -5.57877 & &  -0.68950 & &  -3.300 & &  -2.630 \\
0.70 & &  -8.45564 & &  -3.31269 & &  -0.21241 & &  -3.630 & &  -3.280 \\
0.80 & &  -6.62135 & &  -1.48119 & &   0.05530 & &  -4.300 & &  -3.950 \\
\hline
\hline
\end{tabular}
\end{table}
\end{center}
\begin{center}
\begin{table}[ht]
\caption{\label{coef1.5}
Second-order polynomial fit coefficients for $\lambda=1.5$.
}
\begin{tabular}{ccccccccccc}
\hline
\hline
$\rho^*$ & & $a_0$ & & $a_1$ & & $a_2$ & & $u^*_{min}$ & & $u^*_{max}$ \\
\hline
0.10 & &  -1.65393 & &  -3.97199 & &  -1.56668 & &  -1.100 & &  -0.574 \\
0.20 & &  -2.94547 & &  -3.55764 & &  -0.84737 & &  -1.795 & &  -1.205 \\
0.30 & &  -5.15246 & &  -4.16089 & &  -0.72716 & &  -2.495 & &  -1.900 \\
0.40 & &  -8.96461 & &  -5.41218 & &  -0.74406 & &  -3.145 & &  -2.654 \\
0.50 & & -11.88410 & &  -5.45835 & &  -0.56681 & &  -3.825 & &  -3.400 \\
0.60 & &  -9.17055 & &  -2.83072 & &  -0.14575 & &  -4.555 & &  -4.200 \\
0.70 & &  -7.50855 & &  -1.45483 & &   0.01771 & &  -5.295 & &  -4.950 \\
0.80 & &  -1.37322 & &   1.14105 & &   0.24968 & &  -5.975 & &  -5.650 \\
\hline
\hline
\end{tabular}
\end{table}
\end{center}

\begin{figure}
\begin{center}
\includegraphics[width=7.50cm,clip]{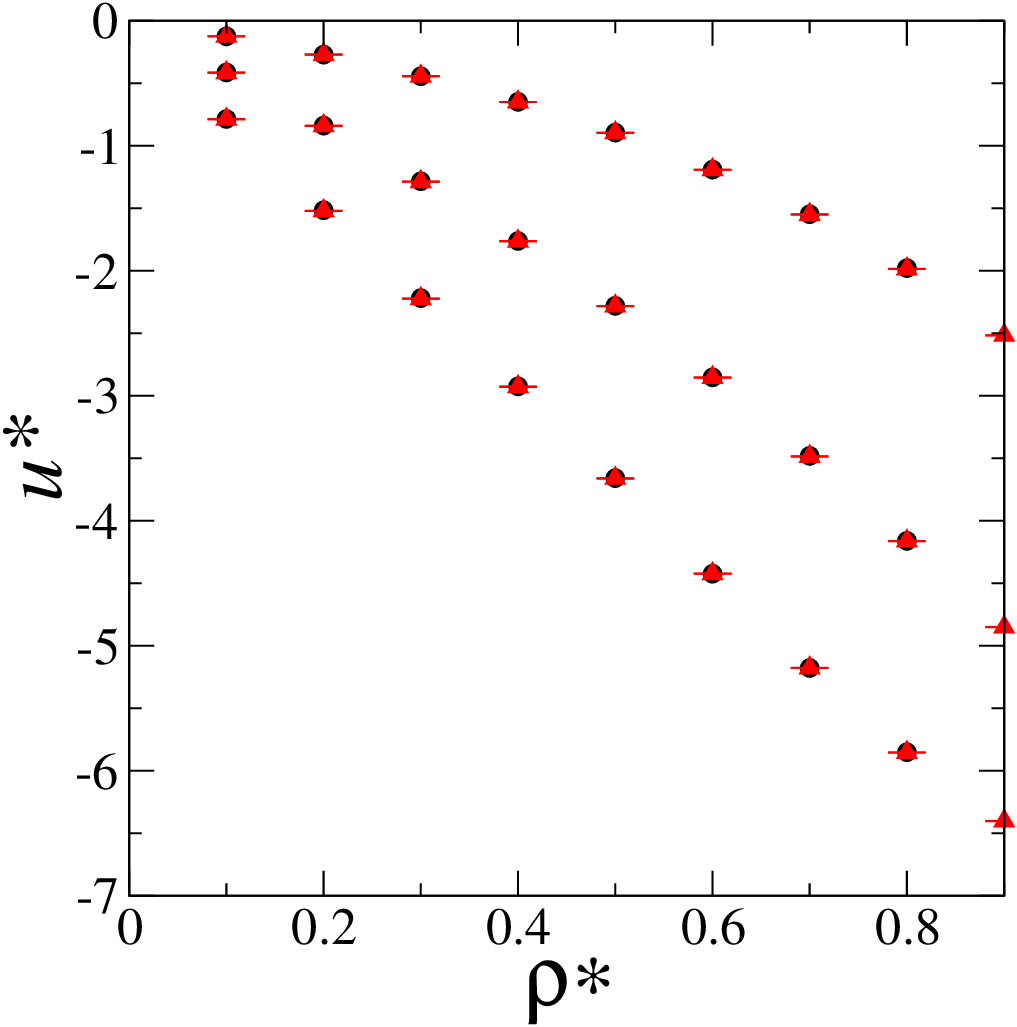}
\caption{\label{energias} (Color on line)
Excess energies of 
SW fluids with attractive ranges $\lambda = 1.1$, 1.3 and 1.5,
for the isotherm $T^*=2.0$ and several densities $\rho^*$.
Black circles correspond to results obtained with the polynomial expressions
with coefficients given in tables I-III, and red triangles are
NVT MC simulation data generated in this work.
}
\end{center}
\end{figure}

From the fitted expressions it is possible then to evaluate the energies and heat capacities,
for a given temperature $T^*$ in the range of validity of the polynomial (\ref{polinomio}).
In Figure \ref{energias} we 
present the energy values for the isotherm $T^*=2.0$; results are
compared with conventional MC-NVT simulated values, obtained 
using 864 particles, with $2.5\times 10^5$ cycles required for equilibration and  $5.0\times 10^5$ cycles
to obtain averaged quantities.

\begin{figure}
\begin{center}
\includegraphics[width=7.50cm,clip]{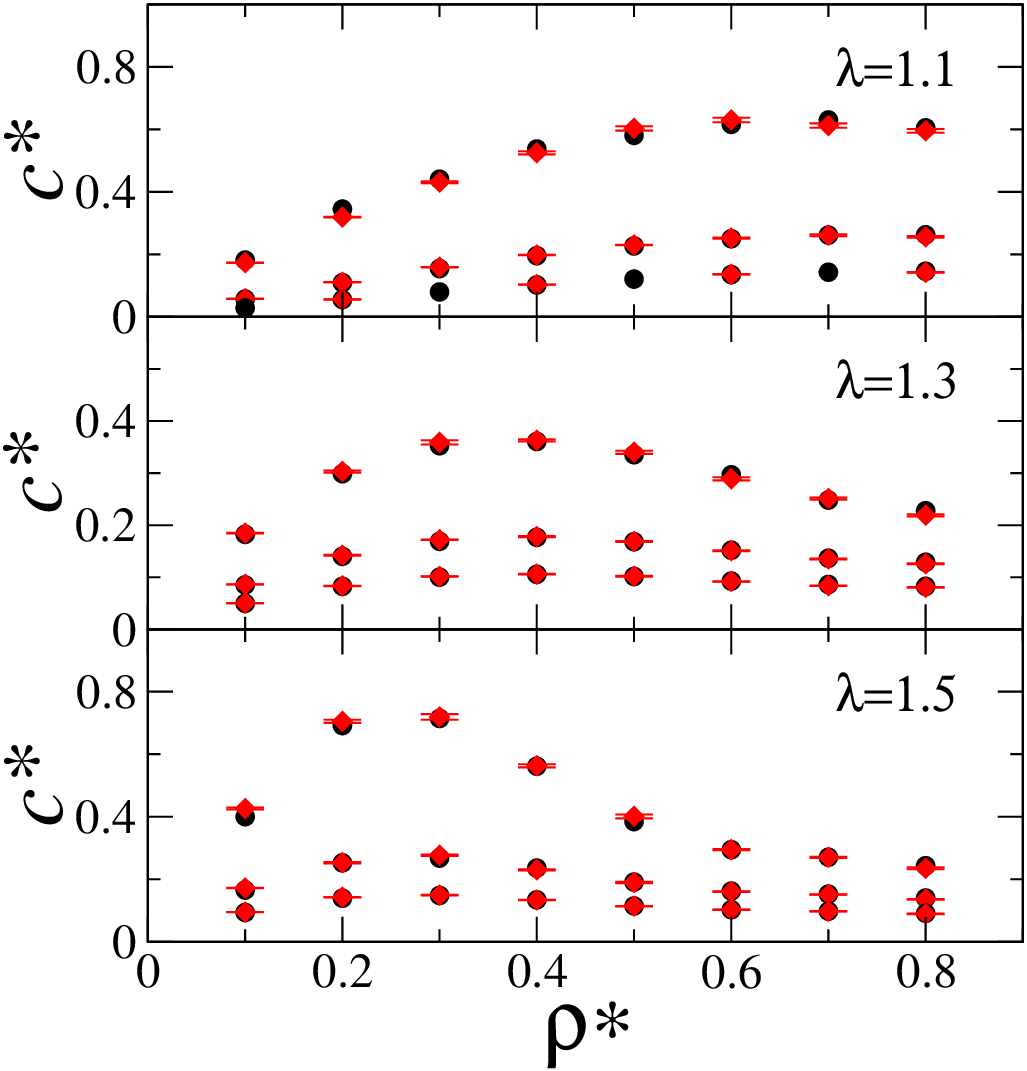}
\caption{\label{capacidad} (Color on line)
Isochoric heat capacities of  
SW fluids with attractive ranges $\lambda = 1.1$, 1.3 and 1.5,
for several temperatures $T^*$ and densities $\rho^*$:
$\lambda=1.1$ with $T^*=1.0$, 1.5 and 2.0 from top to bottom,
$\lambda=1.3$ with $T^*=1.5$, 2.0 and 2.5 from top to bottom, and
$\lambda=1.5$ with $T^*=1.5$, 2.0 and 2.5 from top to bottom.
In all cases black circles correspond to results obtained with the polynomial expressions
whose coefficients are given in tables I-III and red triangles are MC-NVT results
from reference~\cite{Largo2003}.
}
\end{center}
\end{figure}

In Figure \ref{capacidad} results are presented
for the reduced isochoric heat capacity, $c^* = c(E)/N\epsilon$ 
and are compared with MC-NVT values reported by 
Largo {\em et al.}~\cite{Largo2003}, obtaining a remarkable
compatibility between results. 

\section{Conclusions}
In this work 
we have adapted a microcanonical algorithm,
originally developed for the Ising model, to simulate 
thermodynamic properties
of fluids whose molecules
interact via a pair SW potential of variable range. 
The algorithm has the advantage with respect to
a MC-NVT simulation that
a continuous range of temperatures is obtained for a given density,
and it is also possible to predict accurate results for the internal energy and the isochoric heat capacity.
Since the method is based on the Ising model using discrete values,
in the case of SW fluids can only be applied to evaluate entropy derivatives with respect
to discrete quantities, like the number of particles, that would be useful in order to obtain
chemical potentials.
Since the SW fluid is the key ingredient to study other DP systems, we are currently studying
its extension to these generalized models of fluids. Finally, 
to implement  a protocol for
continuous systems could be possible,
considering that 
the probabilities to perform and revert each allowed movement in
the states space of the system are the same,
and defining a probability to obtain an energy 
change starting from a given configuration~\cite{mun:cpc99}. 

\section*{Acknowledgments}
We thank the financial support from CONACYT (M\'exico): Project No. 152684 and Universidad de Guanajuato (M\'exico)
Grant 56-060.  ALB and AGV thank CONACYT (M\'exico) Convocatorias 2014 y 2015 de Estancias Sab\'{a}ticas Nacionales, Estancias
Sab\'{a}ticas al Extranjero y Estancias Cortas para la Consolidaci\'{o}n de Grupos de Investigaci\'{o}n.

\end{document}